\documentclass[preprint2]{aastex}

\usepackage[dvips]{color}

\newcommand{\ME}{M_\oplus}
\newcommand{\sub}[1]{_{\rm #1}}

\newcommand{\pderiv}[2]{\frac{\partial #1}{\partial #2}}

\shorttitle{Accretion of hydrogen-rich atmospheres on super-Earths}
\shortauthors{Ikoma and Hori}

\begin{document}

\title{In-situ Accretion of Hydrogen-Rich Atmospheres on Short-Period Super-Earths: Implications for the Kepler-11 Planets}

\author{M. Ikoma}
\affil{Department of Earth and Planetary Science, The University of Tokyo, 
7-3-1 Hongo, Bunkyo-ku, Tokyo 113-0033, Japan}
\email{ikoma@eps.s.u-tokyo.ac.jp}
\author{Y. Hori}
\affil{Division of Theoretical Astronomy, National Astronomical Observatory of Japan, 2-21-1 Osawa, Mitaka, Tokyo 181-8588, Japan}
\email{yasunori.hori@nao.ac.jp}

\begin{abstract}
Motivated by recent discoveries of low-density super-Earths with short orbital periods, we have investigated in-situ accretion of H-He atmospheres on rocky bodies embedded in dissipating warm disks, by simulating quasi-static evolution of atmospheres that connect to the ambient disk. 
We have found that the atmospheric evolution has two distinctly different outcomes, depending on the rocky body's mass: 
While the atmospheres on massive rocky bodies undergo runaway disk-gas accretion, 
those on light rocky bodies undergo significant erosion during disk dispersal. 
In the atmospheric erosion, the heat content of the rocky body that was previously neglected plays an important role. 
We have also realized that the atmospheric mass is rather sensitive to disk temperature in the mass range of interest in this study. 
Our theory is applied to recently-detected super-Earths orbiting Kepler-11 to examine the possibility that the planets are rock-dominated ones with relatively thick H-He atmospheres. 
The application suggests that the in-situ formation of the relatively thick H-He atmospheres inferred by structure modeling  is possible only under restricted conditions; namely, relatively slow disk dissipation and/or cool environments. 
This study demonstrates that low-density super-Earths provide important clues to understanding of planetary accretion and disk evolution.
\end{abstract}


\section{INTRODUCTION}

An increasing number of low-mass exoplanets 
with masses of less than about 20~$M_\oplus$, 
which are often called super-Earths (SEs)\footnote{
While exoplanets with masses of 10-20~$M_\oplus$ are sometimes called exo-Neptunes instead of super-Earths, we simply call them super-Earths in this paper.
}, 
have been detected recently, thanks to progresses in radial-velocimetry and transit photometry including operation of two space telescopes, Kepler and CoRoT. The size measurement allows us to access the planetary interiors theoretically and thereby estimate the planetary bulk compositions. It has been revealed that there are many low-density SEs that are larger in size than rocky (i.e., iron-silicate) objects of the same masses. As for SEs, 
the variety of possible ingredients obscures the composition estimation; namely, causing degeneracy in composition 
\citep[][]{Valencia+07, Seager+07, Adams+08, Grasset+09, Rogers+Seager10}. 
In particular, low-density SEs are subject to the degeneracy. Sometimes, the degeneracy has a crucial impact on understanding the origin of planets and planetary systems.

A typical example is degeneracy originating from the uncertainty of the presence and mass of H-He atmospheres. The mass-radius relationship for a planet alone is insufficient to distinguish whether the planet is 
a rock-dominated or water-dominated one, 
because the H-He atmosphere makes up the difference in size between rocky and 
water bodies of the same mass. Since transiting SEs in general are orbiting close to their host stars (namely, well inside the snow line), whether they are rocky or water planets would affect considerably understanding of the accretion and migration history of the planets.  An example of that is recently-detected SEs orbiting a Sun-like star named Kepler-11.

The Kepler-11 system is a multiple-planet system that contains at least five low-density SEs \citep{Lissauer+11}. Of them, the densities of Kepler-11d and 11e are as low as 0.9$+0.5\atop-0.3$ $\rm g \, cm^{-3}$ and 0.5$+0.2 \atop -0.2$ $\rm g \, cm^{-3}$, respectively, which are lower than those of 
pure-water planets. 
This suggests the presence of H-He atmospheres on the planets. 
Indeed, the structure modeling by \citet{Lissauer+11} 
suggests a possibility that those two SEs are rock-dominated planets with thick H-He atmospheres that account for 10-20~\% of the planetary masses.
If so, they are a new type of planet that we have never seen in the current solar system.

The H-He atmospheres, if primordial, came from the protoplanetary disk where the planets formed. In the framework of the core accretion model \citep{Hayashi+85}, solid planetary embryos first form, and then collect the ambient disk gas gravitationally to form H-He atmospheres. If this proceeds well before disk dispersal, the disk-gas accretion enters a runaway state at some critical point, which results in forming envelopes of gas giants \citep{Mizuno80, Bodenheimer+Pollack86} which are much more massive than the predicted atmospheres of the Kepler-11 planets. The critical point is when the atmospheric mass is approximately 1/4-1/3 of the protoplanet's total mass \citep{Stevenson82, Wuchterl93}---the fraction being similar to those of the atmospheres of interest.  This means that the inferred masses of the H-He atmospheres of the Kepler-11 planets are close to critical, which motivates us to investigate the possible amounts of H-He gas that proto-SEs with short orbital periods gain from protoplanetary disks.

In this study, we explore the possibility of the \textit{in-situ} accretion of H-He atmospheres. 
As for the origin of H-He atmospheres of short-period SEs, possibilities of Neptune-like planets 
(i.e., water-dominated SEs) that form far out and migrate close to host stars \citep{Rogers+11} and remnants of gas giants whose envelopes were stripped \citep{Nayakshin11} were previously discussed. 
The in-situ accretion of H-He atmospheres on rocky SEs from disks has not been explored by direct simulation of disk gas accretion in the mass and orbital period regimes of transiting SEs. 
The in-situ accumulation of H-He atmospheres would be reasonable in the modern picture of planet formation. Short-period SEs are likely to have migrated to their current locations. The promising mechanism to move them inward is the type-I migration which occurs via planet-disk tidal interaction \citep{Ward86}. The protoplanets formed in this way are packed closely together due to resonance capture, according to recent $N$-body simulations \citep{TP07}. After the migration, the disk dissipates. In many cases, the decline of disk gas density triggers dynamical instability of a multiple-protoplanet system, resulting in orbital crossing and collisions of the protoplanets  \citep[e.g.,][]{Ogihara+Ida09}.  Subsequently, the protoplanets collect the surrounding disk gas to form atmospheres in the dissipating disk. Thus, the competition between the atmospheric accumulation and the disk dissipation may yield intermediate-mass atmospheres like the predicted atmospheres for the Kepler-11 planets.

This paper is organized as follows. In section~\ref{sec:accumulation}, we describe our theoretical model of the atmospheric accumulation with emphasis on effects that we newly incorporate in this study. Then, we show numerical results of the atmospheric growth and the sensitivities of the final atmospheric masses to parameters involved in the model. In section~\ref{sec:discussion}, the masses of the accreted atmospheres that we calculate are compared with those of the atmospheres of Kepler-11 planets inferred by internal-structure modeling. Based on the comparison, we discuss the possibility of the in-situ accretion of the atmospheres. Other possibilities are also discussed. Finally, we conclude this paper in section~\ref{sec:conclusion}. 

\section{ACCUMULATION OF ATMOSPHERE}
\label{sec:accumulation}

\subsection{Model}
\label{sec: model}

We simulate the radially 1D structure and quasi-static evolution of the atmosphere of a protoplanet embedded in a protoplanetary disk. 
The atmosphere's contraction (or expansion) results in its mass gain (or loss). 
The detail of the model is described in \citet{Ikoma+Genda06}. 
The numerical integration is done with the code that we developed and used in \citet{Hori+Ikoma10,Hori+Ikoma11}. 
Below is a brief summary of the model.

The planet consists of a compressive atmosphere with solar elementary abundance on top of an incompressive ''rocky'' body with density of 3.9~$\rm g/cm^{3}$. 
Choice of value of the rocky density and incorporation of rocky small compressibility have tiny impacts on the structure and mass of the atmosphere.  
We integrate a usual set of four equations describing the spherically-symmetric, quasi-hydrostatic structure of a self-gravitating atmosphere, which includes the equations of hydrostatic equilibrium, mass conservation, radiative/convective energy transfer, and energy conservation (i.e., time change in entropy).
The equation of state for the atmospheric gas that we use in this study is from \citet{SCVH95}.

The atmosphere is assumed to be equilibrated with the disk at the smaller of the Bondi and Hill spheres \citep[see][]{Ikoma+Genda06}; namely, the atmospheric density and temperature are equal to those of the ambient disk gas there. 
On the other hand, the bottom of the atmosphere corresponds to the interface between the atmosphere and the rocky body. 
The atmosphere is heated by the underlying rocky body at the bottom. 
In this study, the energy flux from the rocky body is given at the atmospheric bottom as an inner boundary condition, as described below.

In this paper, we consider grain-free atmospheres, unless otherwise noted, 
to investigate upper limits to the masses of the H-He atmospheres that the SEs gain via the in-situ accretion. Thus, we assume that the atmospheric opacity includes only the gas opacity, which is taken from \citet{Freedman+08} in this study. 
We suppose that the protoplanet formed via giant impacts in a dissipating disk. Each simulation of the atmospheric growth starts with an arbitrarily hot state (i.e., a given high luminosity). The initial choices of the luminosity do not affect our results because the initial cooling of the atmosphere occurs fast relative to the disk dissipation. No continuous energy supply by planetesimal bombardment is assumed. 
This assumption also leads to finding upper limits of the atmospheric mass.

In this study, we include two effects newly: We assume that disk dissipation occurs concurrently with the atmospheric growth. Furthermore, we incorporate the effect of the heat capacity of the rocky body. While both effects were not included in previous studies, they have crucial impacts on atmospheric growth in the situations considered in this study, as shown below.

The disk gas density, $\rho_d (a, t)$, is assumed to decrease in an exponential fashion, namely, 
\begin{equation}
	\rho_d (a, t) = \rho_d^0 (a) \exp{(-t/\tau_d)}, 
\end{equation}	
where $a$ is the semi-major axis, $t$ is time, and $\tau_d$ is the dissipation time that is regarded as a parameter; 
the initial disk density, 
$\rho_d^0 (a)$, is taken from \citet{Hayashi81} in this study. According to modern theories of disk evolution \citep[][for a review]{Calvet+00}, the overall dissipation first occurs via viscous diffusion in $\sim 10^7$~yr. Once the gas density becomes so low in the inner disk that the stellar UV penetrates through the disk, photo-evaporation occurs around the gravitational radius of several AU. 
From that time on, the inner disk is separated from the outer disk and evolves via viscous diffusion separately. Since this is equivalent to the fact that the disk size decreases by a factor of 10, the diffusion timescale (being proportional to the square of disk size) for the inner disk decreases by a factor of 100 to become $\sim 10^5$~yr. 
Based on a simple analytical argument, 
namely, solving the 1D viscous-diffusion equation for surface gas density, $\Sigma_d$, 
\begin{equation}
	\pderiv{\Sigma_d}{t} = \frac{1}{a} \pderiv{}{a} \left[3 a^{1/2} \pderiv{}{a} \left(\Sigma_d \nu a^{1/2} \right) \right]
\end{equation}
with $\Sigma_d = 0$ at the inner and outer edges, 
one finds that the inner disk dissipates in such a way that $\rho_d \propto \exp (- \pi^2 t / 10^5 {\rm yr})$, which corresponds to $\tau_d \sim 10^4 {\rm yr}$. This is consistent with results from recent numerical simulations of disk evolution \citep[e.g.,][]{GDH09}.

Decrease in disk density results in cooling of the atmosphere \citep{Ikoma+Genda06}. The rocky body below the atmosphere also cools down. 
Detailed treatment of the rocky body's thermal evolution is beyond the scope of this paper. 
Instead, we evaluate its contribution by inputting the luminosity from the rocky body in the form of $L\sub{rock} = - C\sub{rock} M\sub{rock} {dT_b}/{dt} + L_\mathrm{radio}$ as an inner boundary condition for the atmospheric structure, where $C\sub{rock}$ is the specific heat of silicate ($= 1.2 \times 10^7\, \rm erg \, K^{-1}g^{-1}$), $M\sub{rock}$ the mass of the rocky body, $T_b$ the temperature at the atmospheric bottom, and $L_\mathrm{radio}$ the luminosity due to the radioactive decay of chondrites, $2 \times 10^{20} (M_\mathrm{rock}/M_\oplus)\, \rm erg\, s^{-1}$ \citep[see][]{Guillot+95}. In the present case, since the decline of $T_b$ is caused by that of $\rho_d$,
\begin{equation}
  \frac{dT_b}{dt} = \frac{d \ln \rho_d}{dt} \frac{dT_b}{d \ln \rho_d} 
                  = - \frac{1}{\tau_d} \frac{dT_b}{d \ln \rho_d} .
\end{equation}
Furthermore, from the analytical solution of the radiative atmosphere \citep{Ikoma+Genda06}, $T_b$ is related to the disk temperature, $T_d$, as $dT_b / d \ln \rho_d = T_d /4$. Hence, we calculate $L\sub{rock}$ as
\begin{equation}
  L\sub{rock} = \frac{M\sub{rock} C\sub{rock} T_d}{4 \tau_d} + L_\mathrm{radio}.
\end{equation}

\subsection{Atmospheric growth}

\begin{figure}[tb]
  \begin{center}
  \includegraphics[width=7.5cm,keepaspectratio]{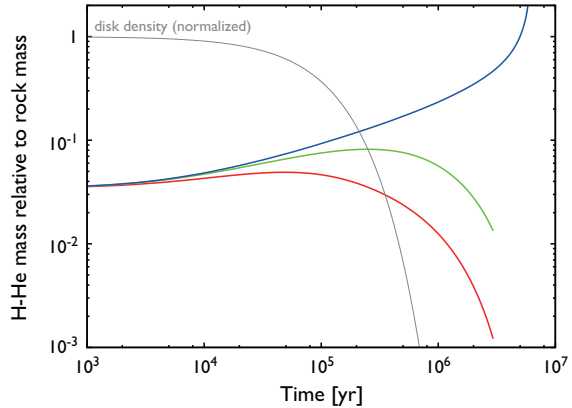}
  \caption{\footnotesize  
Evolution of the atmosphere on the 4$\ME$ rocky body for disk temperature $T_d$ = 550~K. 
In the blue-line case, the disk density, $\rho_d$, is assumed to be constant through the simulation. 
In the green-line and red-line cases, the disk is assumed to dissipate as 
$\rho_d \propto \exp{\left(- t / {10^5 \rm yr} \right)}$, 
which is shown by a thin gray line. 
Also, 
the rocky heat capacity, $C\sub{rock}$, is $1.2 \times 10^7$~$\rm erg \, K^{-1} \, g^{-1}$ in the red-line case, while $C\sub{rock} = 0$ in the green-line case. Note that the radioactive luminosity is also included, but has a negligible impact on the atmospheric evolution.
}
  \label{fig:growth}
  \end{center}
\end{figure}

Examples of atmospheric growth are shown in Figure~\ref{fig:growth}. In the simulations, $M\sub{rock}$ is assumed to be 4~$\ME$. 
The blue line shows the atmospheric growth when the disk gas density is constant through the simulation. In this case, after gradual growth, runaway gas accretion starts at $t \sim$ 3 Myr, resulting in forming an atmosphere that is much more massive than the atmospheres of interest in this study.

The green and red lines show the atmospheric mass evolution with concurrent disk dissipation with $\tau_d = 1 \times 10^5$ years; 
the former and latter represent the cases with $C\sub{rock} = 0$ 
(i.e., no contribution of the rocky body's cooling to heating the atmosphere) 
and $C\sub{rock} = 1.2 \times 10^7\,{\rm erg \, K^{-1}g^{-1}}$, respectively. 
As shown by the two lines, the atmospheric growth levels off at some point in these cases, 
because the disk is assumed to disperse before the runaway gas accretion would happen if it were not for disk dispersal. 
After the leveling-off, the atmosphere is found to be eroded. 
The simulations were stopped when the disk gas density became low enough that the mean-free path of the ambient gas molecules was longer than the planetary radius; 
\begin{eqnarray}
	\rho_d &=& \frac{k T_d}{G M_\mathrm{rock} \sigma} \nonumber \\
		&=& 1.4 \times 10^{-19} \left(\frac{M_\mathrm{rock}}{1\,M_\oplus}\right)^{-1}
		\left(\frac{T_d}{100~\mathrm{K}} \right) ~{\mathrm{g/cm}^3} \\
		&=& 10^{-10} \rho_d^0 (a) \left(\frac{M_\mathrm{rock}}{1\,M_\oplus}\right)^{-1}
		\left(\frac{T_\mathrm{d}}{100\mathrm{K}} \right) \left( \frac{a}{1\mathrm{AU}}\right)^{11/4}, \nonumber 
\end{eqnarray}
where $k$ is the Boltzmann constant, $G$ the gravitational constant, and $\sigma$ the collisional cross-section of the molecule ($= 2.5 \times 10^{-16}\, \mathrm{cm}^2$).

The atmospheric erosion during disk dissipation has been newly found in this study. 
The erosion is due to atmospheric expansion. 
The expansion occurs, because disk depressurization enhances the outward pressure gradient near the outer boundary, which pushes the atmosphere outwards. 
Then, the atmospheric gas seeps out from the protoplanet's gravitational sphere, which continues until the atmosphere is equilibrated with the ambient, depressurized disk. 
The decrease in atmospheric mass due to this effect (the green line) is, however, small, relative to the decrease due to heatup by the cooling rocky body (the red line). 
As described in section~\ref{sec: model}, the disk dissipation cools the surface of the rocky body, resulting in heat supply to the atmosphere. 
This lifts up the atmosphere, which results in further atmospheric erosion.
This is why the atmospheric erosion is more significant 
when the rocky heat capacity is incorporated.

\subsection{Final masses of the H-He atmospheres}
\label{sec: final masses}

\begin{figure}[tb]
  \begin{center}
  \includegraphics[width=7.5cm, keepaspectratio]{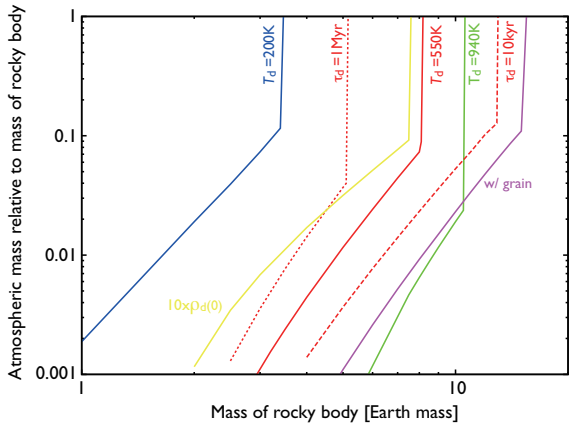}
    \caption{\footnotesize 
The final atmospheric mass as a function of the rocky body's mass. 
The red solid line represents a fiducial model where the disk temperature, $T_d$, is 550~K, the disk dissipation time, $\tau_d$, is 100~kyr, and the initial disk density, $\rho_d^0$, is equal to that of the minimum-mass solar nebula, $\rho\sub{MSN}$ \citep{Hayashi81}; the values of the three parameters are used below unless otherwise specified. 
The red dashed and dotted lines are for $\tau_d =$ 10~kyr and 1~Myr, respectively. 
The green and blue lines are for $T_d$ = 940~K and 200~K, respectively. 
In the yellow-line case, $\rho_d^0$ is 10 times as large as $\rho\sub{MSN}$.
 While grain-free opacity is assumed in the other cases, the grain opacity for protoplanetary disks from \citet{Semenov+03} is used in the purple-line case. 
}
      \label{fig:final mass}
  \end{center}
\end{figure}

In Fig.~\ref{fig:final mass}, the final mass of the H-He atmosphere, $M\sub{H+He}$, which is divided by the rocky body's mass, $M\sub{rock}$, is shown as a function of $M\sub{rock}$ for different values of the disk's temperature, $T_d$, and dissipation time, $\tau_d$. While those two parameters are focused on in this study because upper limits of $M\sub{H+He}$ are of special interest, 
the sensitivity of $M\sub{H+He}$ to the initial disk density, $\rho_d^0$, and the atmospheric opacity is briefly seen below.

First, for a given set of the parameters, $M\sub{H+He}$ increases with $M\sub{rock}$ in the small-$M\sub{rock}$ regime.  Also, it is found that there is a critical value of the rocky body's mass, $M\sub{rock}^\ast$, beyond which the atmosphere becomes quite massive eventually. This is because the atmospheric accretion enters the runaway state before the disk is depleted significantly. Thus, the atmospheric growth has two distinctly different outcomes, depending on $M\sub{rock}$. The atmospheric growth beyond the critical point is discussed in section~\ref{sec:discussion}.

The three red lines of different types show the sensitivity of $M\sub{H+He}$ to $\tau_d$. 
For a given $M\sub{rock}$, larger $\tau_d$ results in larger $M\sub{H+He}$. 
The sensitivity is not large; $M\sub{H+He}$ increases by a factor of less than 10 for two-order-of-magnitude increase in $\tau_d$. 
Also, $M\sub{rock}^\ast$ increases, as $\tau_d$ decreases. This is because a short lifetime of the disk (i.e., small $\tau_d$) allows only massive protoplanets to start runaway accretion.  Difference in $\tau_d$ also 
makes only a small 
change in $M\sub{rock}^\ast$ and also $M\sub{H+He}$ at the critical point ($< 0.1 M\sub{rock}$). 
It is noted that while physically small, the sensitivities of $M\sub{H+He}$ may suffice to validate observationally inferred masses of the atmospheres, as discussed in section~\ref{sec:discussion}.

The results for different three disk temperatures are shown by the solid green ($T_d = 940$~K), red ($T_d = 550$~K), and blue ($T_d = 200$~K) with $\tau_d = 100$~kyr in Fig.~\ref{fig:final mass}. It is found that $T_d$ has a significant impact on $M\sub{H+He}$ and $M\sub{rock}^\ast$. 
Qualitatively, high $T_d$ results in small $M\sub{H+He}$ and, therefore, large $M\sub{rock}^\ast$. 
This is basically because atmospheres made from warmer gas (more exactly, higher-entropy gas) are gravitationally less bound. 
However, the impact of the outer boundary conditions on the atmospheric mass is known to be negligibly small in the case of \textit{massive} atmospheres embedded in \textit{cool} disks like proto-envelopes of gas giants. 
This is because most of the atmospheric mass is concentrated in the deep region of the atmosphere whose structure is insensitive to the outer boundary conditions \citep{Mizuno80, Stevenson82,Wuchterl93,Ikoma+01}. 
In contrast, in the case of \textit{low-mass} atmospheres embedded in \textit{warm} disks which are appropriate to SEs with short orbital periods, our results in Fig.~\ref{fig:final mass} reveal that the impact is significantly large.

The initial disk density, $\rho_d^0$, has only a small impact on $M\sub{H+He}$ and $M\sub{rock}^\ast$.
For example, shown by the yellow line is the result for the case of $\rho_d^0$ being 10 times larger than that used for the red solid line. Since high density corresponds to low entropy, the atmosphere embedded in denser disks tends to be more massive for the same reason described above. However, it is realized that the atmosphere before being eroded is massive enough that the structure in its deep part is insensitive to disk properties.

Finally, 
the values of $M\sub{H+He}$ and $M\sub{rock}^\ast$ that we have derived are upper and lower limits, respectively, for a given set of $T_d$ and $\tau_d$ in the sense that no grain opacity and no additional energy input are assumed. 
In the case represented by the purple line, the effect of grain opacity is evaluated. 
To do so, we have adopted \citet{Semenov+03}'s protoplanetary-disk grain-opacity model. This would be an extreme case of large opacity, namely, a lower limit to $M\sub{H+He}$ because large opacity slows down contraction of the atmosphere \citep{Ikoma+00}. Thus, the actual solutions are in between the lines of grain-free (red solid line) and grain-rich (purple line) atmospheres.

\section{APPLICATION TO KEPLER-11 SUPER-EARTHS}
\label{sec:discussion}
\subsection{Comparison with atmospheric masses inferred by modeling}

\begin{figure}[tb]
  \begin{center}
  \includegraphics[width=7.5cm, keepaspectratio]{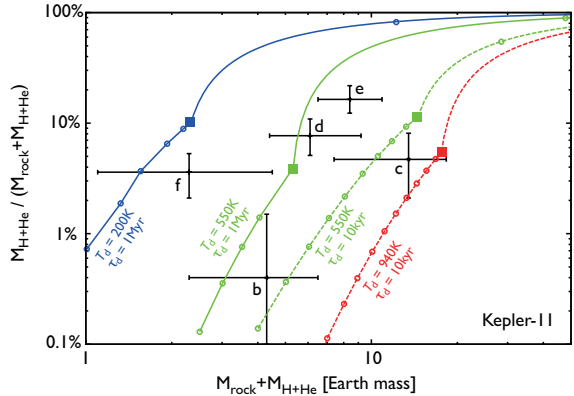}
    \caption{\footnotesize  
Application to the Kepler-11 planets.
The black crosses represent the estimated atmospheric masses of Kepler-11b to 11f, 
the data of which were provided by E. Lopez (personal communication). 
The red, green, and blue lines are for $T_d =$ 940~K, 550~K, and 200~K, respectively. 
The solid and dashed lines are for $\tau_d = 1$~Myr and $10$~kyr, respectively. 
We have drawn the extrapolated lines beyond the critical points represented by filled squares, assuming that H-He gas accumulates on the rocky body with the critical mass. 
The points on the extrapolation lines are the results of simulations where the atmospheric accretion is limited by supply of gas from the disk that evolves via viscous diffusion; $\alpha$-viscosity is adopted and the values of $\alpha$ are $1 \times 10^{-4}$ for $\tau_d = 1$~Myr and  $1 \times 10^{-2}$ for $\tau_d = 10$~kyr.
}
      \label{fig:comparison}
  \end{center}
\end{figure}

The super-Earths orbiting Kepler-11 are thought to have relatively thick H-He atmospheres, as described in Introduction. In this section, we apply our theory to the planets and, thereby, get some implications for their composition and origin. Specifically, we compare the atmospheric masses derived from modeling with those which we have calculated in this study.

The comparison is made in Fig.~\ref{fig:comparison}: The $x$-axis is the planetary total mass, $M_p$ (i.e., $M\sub{rock}+M\sub{H+He}$); the $y$-axis is the percentage of $M\sub{H+He}$ relative to $M_p$.  The black crosses represent $M\sub{H+He}$ derived by the structure modeling of the planets, which account for the observed masses and radii. The data were kindly provided by E. Lopez who calculated them using a modified version of the modeling method from \citet{Lissauer+11}; the effect of the rocky heat capacity having been incorporated in the thermal evolution, which yielded a slight decrease in $M\sub{H+He}$ (Lopez, personal communication). 

The curves show the results of our calculations in this study for four different cases. The temperatures 940~K  and 550~K correspond to those at Kepler-11b's and 11f's semimajor axes ($a = 0.088$ and $0.25$AU), respectively, in an optically-thin disk \citep{Hayashi81}. Points indicated by filled squares are the critical points (see section~\ref{sec: final masses}). 
Beyond the critical points, we have also drawn extrapolated lines (hereafter called super-critical lines) by assuming that H-He atmospheres accrete on the 
non-growing rocky bodies with the critical masses; 
mathematically, the super-critical lines are expressed by 
$y = (1 - M\sub{rock}^\ast / x) \times 100\% $, 
where $x = M\sub{rock}+M\sub{H+He}$ and $y = M\sub{H+He}/(M\sub{rock}+M\sub{H+He})$. 
To check the validity of the super-critical lines, we have simulated the super-critical (i.e., runaway) accretion whose rate is not higher than supply limit of the disk gas due to viscous diffusion, $3 \pi \Sigma_d \nu_d$, where $\Sigma_d$ and $\nu_d$ are the surface density and viscosity of the disk gas, respectively. We have adopted the $\alpha$-prescription for the disk viscosity; $\tau_d$ = 10~kyr and 1~Myr correspond to $\alpha \simeq 10^{-2}$ and $10^{-4}$, respectively. 
It is shown that the numerical solutions are certainly on the super-critical lines. In the case of $T_d$ = 550~K and $\tau_d$ = 10~kyr, for example, $M\sub{rock} = 13.0M_\oplus$ at the point on the super-critical line, which is close to $M\sub{rock}^\ast$ = 12.9~$\ME$; nevertheless, $M\sub{H+He}/M_p$ is larger by a factor of $\sim$5. This demonstrates that a slight difference in $M\sub{rock}$ around $M\sub{rock}^\ast$ results in a large difference in $M\sub{H+He}$.

Recent $N$-body simulations of the formation of hot SEs \citep{TP07, Ogihara+Ida09} suggest that disk dissipation often causes collisions between embryos of SEs. The major accretion of atmospheres occurs after the collisions. In the late stages of disk evolution, photo-evaporation separates the inner disk from the outer disk; then, the inner disk dissipates on a timescale of as short as $10^4$ years \citep{GDH09}.  The dashed lines in Fig.~\ref{fig:comparison} correspond to the cases of $\tau_d$ = $1 \times 10^4$~yr. Comparing the lines with the crosses, one finds that the values of $M\sub{H+He}$ inferred from modeling are much larger than those derived in this study, except for Kepler-11b and 11c. Furthermore, because the atmospheres are assumed to contain no grains, the derived atmospheric masses are maximal. Thus, the predicted atmospheres of the Kepler-11 planets are not accounted for by the in-situ accretion in such a quickly dissipating disk.

Slower disk dissipation may be preferable in this respect. The green solid line shows the case of $\tau_d$ = 1~Myr and $T_d$ = 550~K. While the corresponding line for $T_d$ = 940~K is not drawn, the difference between the solid and dashed lines is similar to that for $T_d$ = 550~K. In this case, as far as Kepler-11b to 11e are concerned, the values of $M\sub{H+He}$ inferred from modeling seem to be consistent with those derived in this study. Kepler-11f lies above the line; only the high-mass end is close to the line. 
Furthermore, a cooler disk would be compatible with the $M\sub{H+He}$ inferred for all the planets. As a reference, we show the result for $T_d$ = 200~K by a blue solid line. However, it would be worth noting that Kepler-11d and 11e are near the critical points in the super-critical regime. As seen above, such cases are rare, because slightly large cores result in much more massive $M\sub{H+He}$.

\subsection{Mass Loss}

\begin{figure}[tb]
  \begin{center}
  \includegraphics[width=7.5cm,  keepaspectratio]{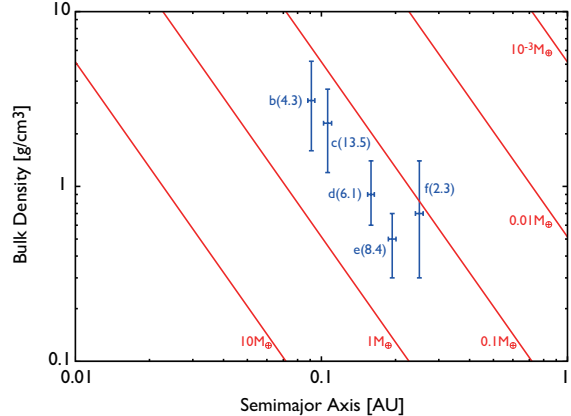}
  \caption{\footnotesize  
      The mass of atmospheric gas that a planet loses
      for 10 Gyr (solid lines). The numbers attached to lines indicate
      the estimated values.  The mass loss is assumed to occur via the
      energy-limited hydrodynamic escape driven by XUV from a G dwarf
      like Kepler-11. Stellar-XUV data have been taken from
      \citet{Ribas+05}.  Measured semimajor axes and bulk densities of
      Kepler-11b, 11c, 11d, 11e, and 11f \citep{Lissauer+11} are shown
      by blue crosses. The numbers in parentheses are
      the likelihood values of the measured planetary masses. }
  \label{fig:lost mass}
  \end{center}
\end{figure}

Before concluding this study, we discuss the loss of the atmospheres. The Kepler-11 planets are orbiting relatively close to their host star which is a G dwarf aged $8\pm2$~Gyr. Since the star is old, the current irradiation level of X-ray and EUV (XUV) is low.  In the past, however, the planets should have been exposed to intense stellar XUV. Stellar-XUV irradiation results in loss of planetary atmospheres.

Here we estimate the mass of the H-He atmosphere that a planet with a given density loses via the energy-limited hydrodynamic escape \citep[e.g.,][]{Watson+81}.  We integrate the equation
\begin{equation}
     \dot{M}\sub{loss} = \frac{3 \epsilon F\sub{XUV}}
                              {4 G \bar{\rho} K\sub{tide}},
\end{equation}
where $\epsilon$ is the heating efficiency, $F\sub{XUV}$ the incident flux of stellar XUV, $\bar{\rho}$ the planetary bulk density, and $K\sub{tide}$ a correction factor for the Roche-lobe effect \citep{Erkaev+07, Lecavelier+04}.  We adopt $\epsilon = 0.4$ \citep[see][]{Valencia+10}, the fitting formula of $F\sub{XUV}$ from \citet{Ribas+05}, and $K\sub{tide}$ = 0.92 appropriate for the Kepler-11 system.

Figure~\ref{fig:lost mass} shows the mass that the planet loses for 10~Gyr as a function of semimajor axis and planetary bulk density, together with the measured values of $a$ and $\bar{\rho}$ for Kepler-11b to 11f. It is revealed that despite of different distances to the host star, all the planets have lost similar amounts of atmospheric gas, which are of the order of 0.1~$\ME$. 
Even for Kepler-11b ($M_p = 4.3 {+2.2 \atop -2.0} \ME$) and 11f ($M_p = 2.3 \pm 1.2 \ME$) which are the two lightest planets among them, the amounts of mass loss correspond to $\sim$1-10~\% of the planetary masses. (Although their early-stage inflated structure being neglected in this estimate, the results are similar to those from more detailed simulations by Lopez et al.~[in preparation]). Thus, the consideration of mass loss does not affect the implications obtained above for Kepler-11c to 11f. As for Kepler-11b, the condition becomes more severe.

\subsection{Other Possibilities}
\subsubsection{Degassing}
Hydrogen-rich atmospheres may also be formed on rocky super-Earths via degassing. Planetesimals contain metallic iron, which is oxidized by water to produce hydrogen \citep{Abe+00}. However, detailed calculations show that the resultant hydrogen-rich atmosphere accounts for, at most, several percents of the planetary mass \citep{Elkins-Tanton+08, Rogers+11}. This suggests that degassing is insufficient to explain, at least, the inferred masses of the atmospheres of Kepler-11d and 11e.

\subsubsection{Water-dominated planets}
Another possibility is that those planets are water-dominated ones covered with relatively thin H-He atmospheres, namely, warm Neptunes, which would have formed beyond the snow line, followed by inward migration. The structure and formation of warm Neptunes were recently discussed by \citet{Rogers+11}. Looking at results of their modeling, one finds out that the atmosphere constituting $<$~1~\% of the planetary mass is compatible with the observed radii of Kepler-11d and 11e. Such atmospheres are stable for the current level of stellar XUV, as mentioned above \citep[see also][]{Rogers+11}. However, the origin is yet to be explored. Also, the resultant masses and radii of the planets formed in the simulations by \citet{Rogers+11} do not match those of the Kepler-11 planets, although they do not have a special focus on Kepler-11 in the paper. Since migration requires the persistence of the disk gas, further gas accretion should occur after arriving near the host star. Hence the origin of such atmospheres is also a challenging issue.

\section{CONCLUSION}
\label{sec:conclusion}

Motivated by recent discoveries of low-density super-Earths (SEs) with short orbital periods, we have investigated the in-situ accretion of H-He atmospheres on short-period SEs. Specifically, we have simulated the quasi-static evolution of the H-He atmospheres embedded in dissipating, warm disks and, thereby, derived the atmospheric masses at the time when disk gas has disappeared. 
We have also applied our theory to recently-detected short-period SEs orbiting Kepler-11, and examined the possibility that the planets are rocky ones with relatively thick H-He atmospheres. The application demonstrates that the in-situ formation of relatively thick H-He atmospheres inferred by structure modeling  is possible only under restricted conditions; namely, relatively slow disk dissipation and/or cool environments. 

Through the application of our theory to detected low-density SEs, we have gotten some important suggestions for future improvement in the theory of the formation of H-He atmospheres of low-mass planets: 
(1) Atmospheric erosion occurs during disk dissipation. This suggests that not only when but also how protoplanetary disks dissipate are crucial factors for determining the atmospheric mass. (2) The thermal contribution of the rocky body to the atmospheric evolution has a significant impact on the atmospheric erosion. Thus, it is important to investigate in more detail the thermal evolution of the rocky body that concurrently occurs with the accretion of the surrounding atmosphere. 
(3) In relatively hot environments considered in this study, the atmospheric mass is rather sensitive to disk temperature, which means that the late-stage thermal evolution of disks should be also taken into account in simulating the formation of H-He atmospheres of disk origin.

Degeneracy in composition between rocky planets with thick hydrogen-rich atmospheres and water planets is a critical issue in planet formation theories. 
Findings from this study are applicable to any short-period super-Earths and would be helpful for removing the degeneracy. 
As demonstrated in this paper, the atmospheric mass inferred by structure modeling with observed mass and radius can be 
evaluated by comparison with that derived from the accretion theory. 
Also, low-density SEs provide important clues to understanding of planetary accretion and disk evolution. 
Planetary candidates observed by \textit{Kepler} contain about thousand SE-size objects \citep{Borucki+11}.  We expect that follow-up observations will identify them and determine their masses, which will lead to better understanding of the origins of low-mass planets.

\acknowledgments
We are grateful to T. Guillot and J. Fortney for fruitful discussions. 
We also thank E. Lopez for kindly providing us with the data of his modeling of the structure and thermal evolution of the Kepler-11 planets. 
We appreciate the anonymous referee's critical and constructive comments which helped us to improve this paper.
M.~I. is supported by Challenging Research Award from Tokyo Institute of Technology and 
Grant-in-Aid for Scientific Research on Innovative Areas (No.~23103005) from the Ministry of Education, Culture, Sports, Science and Technology (MEXT) of Japan.
Y.~H. is supported by Grant-in-Aid for JSPS Fellows (No.~23003491) from MEXT, Japan.


\begin{thebibliography}{}
\bibitem[Abe et al.(2000)]{Abe+00} Abe, Y., Ohtani, E., Okuchi, T., Righter, K., \& Drake, M.\ 2000, Origin of the Earth and Moon, 413
\bibitem[Adams et al.(2008)]{Adams+08} Adams, E.~R., Seager, S., \& Elkins-Tanton, L.\ 2008, \apj, 673, 1160 
\bibitem[Bodenheimer \& Pollack(1986)]{Bodenheimer+Pollack86} Bodenheimer, P., \& Pollack, J. B. 1986, \icarus, 67, 391
\bibitem[Borucki et al.(2011)]{Borucki+11} Borucki, W.~J., Koch, D.~G., Basri, G., et al.\ 2011, \apj, 736, 19
\bibitem[Calvet et al.(2000)]{Calvet+00} Calvet, N., Hartmann, L., \& Strom, S.~E.\ 2000, Protostars and Planets IV, 377
\bibitem[Cochran et al.(2011)]{Cochran+11} Cochran, W.~D., 
Fabrycky, D.~C., Torres, G., et al.\ 2011, \apjs, 197, 7
\bibitem[Elkins-Tanton \& Seager(2008)]{Elkins-Tanton+08} Elkins-Tanton, L.~T., \& Seager, S.\ 2008, \apj, 685, 1237
\bibitem[Erkaev et al.(2007)]{Erkaev+07} Erkaev, N.~V., Kulikov, Y.~N., Lammer, H., Selsis, F., Langmayr, D., Jaritz, G.~F., \& Biernat, H.~K.\ 2007, \aap, 472, 329
\bibitem[Freedman et al.(2008)]{Freedman+08} Freedman, R.~S., Marley, M.~S., \& Lodders, K.\ 2008, \apjs, 174, 504 
\bibitem[Gorti et al.(2009)]{GDH09} Gorti, U., Dullemond, C.~P., \& Hollenbach, D.\ 2009, \apj, 705, 1237 
\bibitem[Grasset et al.(2009)]{Grasset+09} Grasset, O., Schneider, J., \& Sotin, C.\ 2009, \apj, 693, 722
\bibitem[Guillot et al.(1995)]{Guillot+95} Guillot, T., Chabrier, 
G., Gautier, D., \& Morel, P.\ 1995, \apj, 450, 463
\bibitem[Hayashi(1981)]{Hayashi81} Hayashi, C.\ 1981, Progress of Theoretical Physics Supplement, 70, 35 
\bibitem[Hayashi et al.(1985)]{Hayashi+85} Hayashi, C., Nakazawa, K., \& Nakagawa, Y.\ 1985, Protostars and Planets II, 1100
\bibitem[Hori \& Ikoma(2010)]{Hori+Ikoma10} Hori, Y., \& Ikoma, M.\ 2010, \apj, 714, 1343
\bibitem[Hori \& Ikoma(2011)]{Hori+Ikoma11} Hori, Y., \& Ikoma, M.\ 2011, \mnras, 416, 1419
\bibitem[Ikoma \& Genda(2006)]{Ikoma+Genda06} Ikoma, M., \& Genda, H. 2006, \apj, 648, 696
\bibitem[Ikoma et al.(2000)]{Ikoma+00} Ikoma, M., Nakazawa, K., \& Emori, H. 2000, \apj, 537, 1013
\bibitem[Ikoma et al.(2001)]{Ikoma+01} Ikoma, M., Emori, H., 
\& Nakazawa, K.\ 2001, \apj, 553, 999
\bibitem[Lecavelier des Etangs et al.(2004)]{Lecavelier+04} Lecavelier des Etangs, A., Vidal-Madjar, A., McConnell, J.~C., \& H{\'e}brard, G.\ 2004, \aap, 418, L1 
\bibitem[Lissauer et al.(2011)]{Lissauer+11} Lissauer, J.~J., et al.\ 2011, \nat, 470, 53
\bibitem[Mizuno(1980)]{Mizuno80} Mizuno, H. 1980, Prog. Theor. Phys., 64, 544
\bibitem[Nayakshin(2011)]{Nayakshin11} Nayakshin, S.\ 2011, \mnras, 1274
\bibitem[Ogihara \& Ida(2009)]{Ogihara+Ida09} Ogihara, M., \& Ida, S.\ 2009, \apj, 699, 824
\bibitem[Rogers et al.(2011)]{Rogers+11} Rogers, L.~A., Bodenheimer, P., Lissauer, J.~J., \& Seager, S.\ 2011, \apj, 738, 59
\bibitem[Rogers \& Seager(2010)]{Rogers+Seager10} Rogers, L.~A., \& Seager, S.\ 2010, \apj, 712, 974
\bibitem[Ribas et al.(2005)]{Ribas+05} Ribas, I., Guinan, E.~F., G{\"u}del, M., \& Audard, M.\ 2005, \apj, 622, 680
\bibitem[Saumon et al.(1995)]{SCVH95} 
Saumon, D., Chabrier, G., \& van Horn, H.~M.\ 1995, \apjs, 99, 713
\bibitem[Semenov et al.(2003)]{Semenov+03} Semenov, D., Henning, T., Helling, C., Ilgner, M., \& Sedlmayr, E.\ 2003, \aap, 410, 611
\bibitem[Seager et al.(2007)]{Seager+07} Seager, S., Kuchner, M., Hier-Majumder, C.~A., \& Militzer, B.\ 2007, \apj, 669, 1279 
\bibitem[Stevenson(1982)]{Stevenson82} Stevenson, D. J. 1982, Planet. Space Sci., 30, 755
\bibitem[Tanigawa \& Ikoma(2007)]{TI07} Tanigawa, T., \& Ikoma, M.\ 2007, \apj, 667, 557
\bibitem[Terquem \& Papaloizou(2007)]{TP07} Terquem, C., \& Papaloizou, J.~C.~B.\ 2007, \apj, 654, 1110
\bibitem[Valencia et al.(2007)]{Valencia+07} Valencia, D., Sasselov, D.~D., \& O'Connell, R.~J.\ 2007, \apj, 665, 1413
\bibitem[Valencia et al.(2010)]{Valencia+10} Valencia, D., Ikoma, M., Guillot, T., \& Nettelmann, N.\ 2010, \aap, 516, A20
\bibitem[Ward(1986)]{Ward86} Ward, W.~R.\ 1986, \icarus, 67, 164
\bibitem[Watson et al.(1981)]{Watson+81} Watson, A.~J., Donahue, T.~M., \& Walker, J.~C.~G.\ 1981, \icarus, 48, 150
\bibitem[Wuchterl(1993)]{Wuchterl93} Wuchterl, G.\ 1993, \icarus, 106, 323
\end{thebibliography}
\end{document}